\documentclass[10pt,letterpaper]{article}
\usepackage[top=0.85in,left=2.75in,footskip=0.75in]{geometry}

\usepackage{changepage}

\usepackage[utf8]{inputenc}

\usepackage{textcomp,marvosym}

\usepackage{fixltx2e}

\usepackage{amsmath,amssymb}

\usepackage{cite}

\usepackage{nameref,hyperref}

\usepackage[right]{lineno}

\usepackage{microtype}
\DisableLigatures[f]{encoding = *, family = * }

\usepackage{rotating}

\raggedright
\usepackage[aboveskip=1pt,labelfont=bf,labelsep=period,justification=raggedright,singlelinecheck=off]{caption}

\makeatletter
\renewcommand{\@biblabel}[1]{\quad#1.}
\makeatother

\date{}

\usepackage{lastpage,fancyhdr,graphicx}
\usepackage{epstopdf}
\fancyheadoffset[L]{2.25in}
\fancyfootoffset[L]{2.25in}
\begin{document} 
\vspace*{0.35in}

\begin{flushleft}
{\Large
\textbf\newline{\bf Measure of Node Similarity in Multilayer Networks}
}
\newline
\\
Anders Mollgaard\textsuperscript{1},\\
Ingo Zettler\textsuperscript{2},\\
Jesper Dammeyer\textsuperscript{2},\\
Mogens H.~Jensen\textsuperscript{1},\\
Sune Lehmann\textsuperscript{3},\\
Joachim Mathiesen\textsuperscript{1,*}\\
\bigskip
\bf{1} University of Copenhagen, Niels Bohr Institute , 2100 Copenhagen, Denmark\\
\bf{2} University of Copenhagen, Department of Psychology, 1353 Copenhagen, Denmark\\
\bf{3} Technical University of Denmark, 2800 Kgs.~Lyngby, Denmark\\
\bigskip
* mathies@nbi.dk

\end{flushleft}

\section*{Abstract}
The weight of links in a network is often related to the similarity of the nodes.  Here, we introduce a simple tunable measure for analysing the similarity of nodes across different link weights. 
In particular, we use the measure to analyze homophily in a group of 659 freshman students at a large university. Our analysis is based on data obtained using smartphones equipped with custom data collection software, complemented by questionnaire-based data. The network of social contacts is represented as a weighted multilayer network constructed from different channels of telecommunication as well as data on face-to-face contacts. We find that even strongly connected individuals are not more similar with respect to basic personality traits than randomly chosen pairs of individuals. In contrast, several socio-demographics variables have a significant degree of similarity. We further observe that similarity might be present in one layer of the multilayer network and simultaneously be absent in the other layers. For a variable such as gender, our measure reveals a transition from similarity between nodes connected with links of relatively low weight to dis-similarity for the nodes connected by the strongest links. 
We finally analyze the overlap between layers in the network for different levels of acquaintanceships.

\section*{Introduction}
Are two connected individuals more similar than a pair of strangers? Over the last decades, advances in data collection methods have provided new opportunities for research on human behavior~\cite{borgatti2009network} including the topic of homophily, i.e.,~whether a pair of connected individuals tends to be more similar than pairs of randomly selected individuals. For instance, it is now possible to observe social interaction across multiple channels, e.g.,~by combining data describing face-to-face contacts, with data from online social organizations or smartphone data \cite{eagle2006reality, 10.1371/journal.pone.0011596, aharony2011social, 10.1371/journal.pone.0095978}. Multiple networks formed from the simultaneous interaction in different channels are often called \emph{multiplex} or \emph{multilayer} networks \cite{kivela2014multilayer}. 
Homophily has been observed with regard to many different variables. Examples span across socio-demographic variables (e.g., age, gender, ethnicity), variables describing behavioral patterns (e.g., drinking behavior, smoking behavior, physical activity), variables representing attitudes, beliefs, or opinions (e.g., about politics and sport), and personality traits such as extraversion \cite{Crandall1997133, currarini2010identifying, Kandel1978306,Rushton2005555,kurtz2003relationship, stehle2013gender}. It is an open question though, if homophily is becoming more pronounced between stronger connected individuals. Here, we introduce an extended similarity measure with a tunable parameter, which allows us to check for homophily across links with a broad spectrum of weights. Based on the measure, we find a moderate degree of homophily with respect to behavioral patterns but no significant homophily with regard to the basic personality traits conscientiousness, agreeableness, and neuroticism. 

Most commonly, homophily is investigated via likeability ratings about strangers, via a comparison of personality reports from a dyad, triplet etc.~of acquaintances, or via network analyses. 
Recent studies based on personality reports by well-acquainted persons did find overlap between acquaintances concerning the levels of some of the basic personality traits \cite{Cohen2013816,Lee2009460,Paunonen2013800}. Network studies focusing on observable variables such as gender or cigarette use have suggested that similarity in this regard is important for friendly acquaintanceship \cite{Delay2013464}. Overall, research so far suggests similarity between pairs of friends or acquaintances, but the detailed conclusions concerning homophily tend to differ depending on the methodology. In addition, the similarity of nodes, as we shall see below, is strongly related to the strength of the link connecting them.


For an accurate understanding of homophily, a long-term and detailed monitoring of social networks is needed  for several reasons. In order to reveal a complete picture of homophily, it is essential to gain insights into the similarity at all levels, e.g., from best friends, acquaintances, to people in the network one hardly likes or spends time with. These distinctions are possible in weighted network analyses. 
\begin{figure*}
	\centering
	\includegraphics[width=1\textwidth]{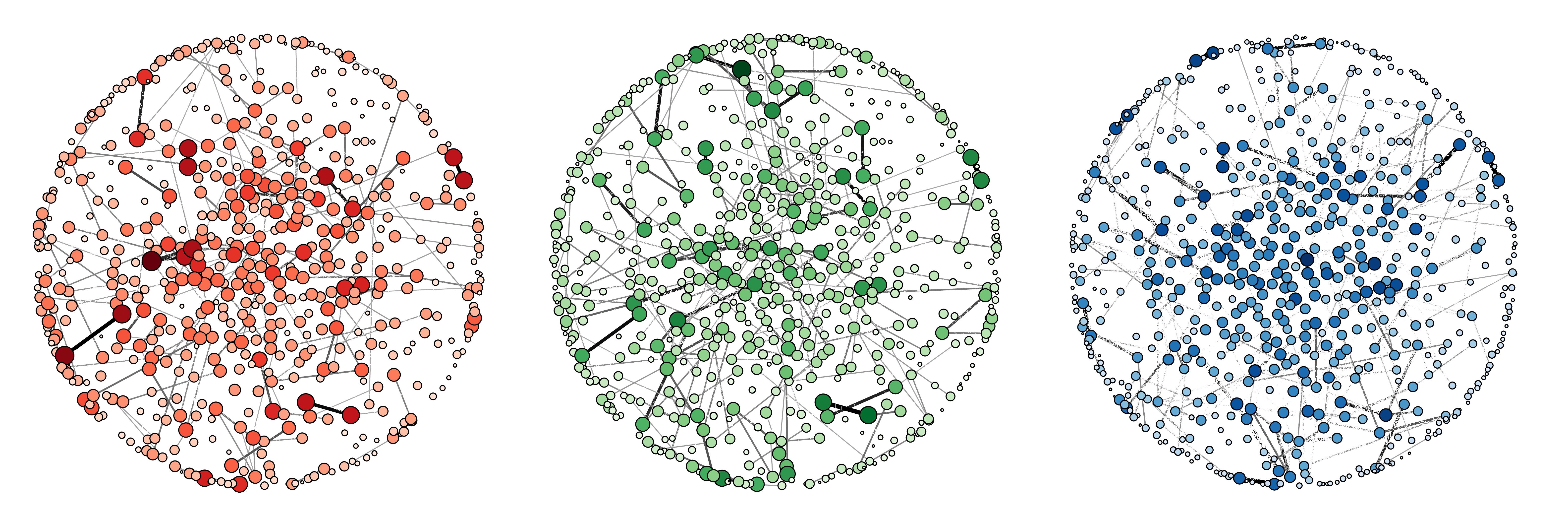}
	\caption{{\bf The phone call, the text message, and the proximity networks.} The size of a node is determined by the sum of the weights of in-going links. The width and the darkness of a link is given by the square root of the link weight. For visual clarity, we show the nodes in the same positions in each panel and we do not show the weakest links ($w_{ij} < 0.01$). \label{fig:networks}}
\end{figure*}
Here, we investigate the similarity of connected individuals in a multilayer social network, with connections based on phone calls, text messages, and physical proximity (Bluetooth). We estimate the similarity between connected persons within a specified network with regard to socio-demographic variables (sex, age, body mass index), behavioral patterns (physical activity, alcohol drinking, and smoking behavior), attitudes concerning politics and religion, and, ultimately, basic personality traits in terms of the \emph{Big Five}, i.e., conscientiousness (e.g., being organized, precise, thorough), agreeableness (e.g., being kind, sympathetic, warm), neuroticism (e.g., being anxious, moody, touchy), openness to experience (e.g., being creative, philosophical, unconventional), and extraversion (e.g., being active, sociable, talkative). We focus on the Big Five as personality traits since they reflect an `integrative descriptive taxonomy for personality research' \cite{john2008paradigm}.

\section*{Results}
This work rests on a unique dataset. We have mapped out the social network between $659$ freshman students starting in the year 2013 at the Technical University of Denmark and running over 24 months \cite{10.1371/journal.pone.0095978}. Using state-of-the-art smartphones equipped with custom data collection software, we have collected the communication patterns within this densely connected population across a number of channels \cite{mollgaard2016dynamics}. Specifically, we measure telecommunication networks (phone calls, text messages), online social networks (Facebook connections and interactions), and networks based on physical proximity. The physical proximity networks are measured via the Bluetooth signal strength, and can be used as a proxy for face-to-face meetings~\cite{sekara2014strength}. As a complement to the network data, we also collect information on geo-spatial mobility using GPS, as well as a number of more technical probes.

In addition to the automated data collection, we have also acquired extensive questionnaire-based data on participants' personality and behavior, comprising the following questionnaires: Big Five Inventory~\cite{john2008paradigm}, Rosenberg Self Esteem Scale~\cite{rosenberg1989society}, Narcissistic Admiration and Rivalry Questionnaire~\cite{back2013narcissistic}, Satisfaction With Life Scale~\cite{diener1985satisfaction}, Rotter's Locus of Control Scale~\cite{rotter1966generalized}, UCLA Loneliness scale~\cite{russell1996ucla}, Self-efficacy~\cite{sherer1982self}, Perceived Stress Scale~\cite{cohen1983global}, Major Depression Inventory~\cite{bech2001sensitivity}, The Copenhagen Social Relation Questionnaire~\cite{lund2014content}, and Positive and Negative Affect Schedule~\cite{watson1988development}, as well as several general health-, attitudes- and behavior-related questions.

\begin{figure}
	\centering
	\includegraphics[width=.5\textwidth]{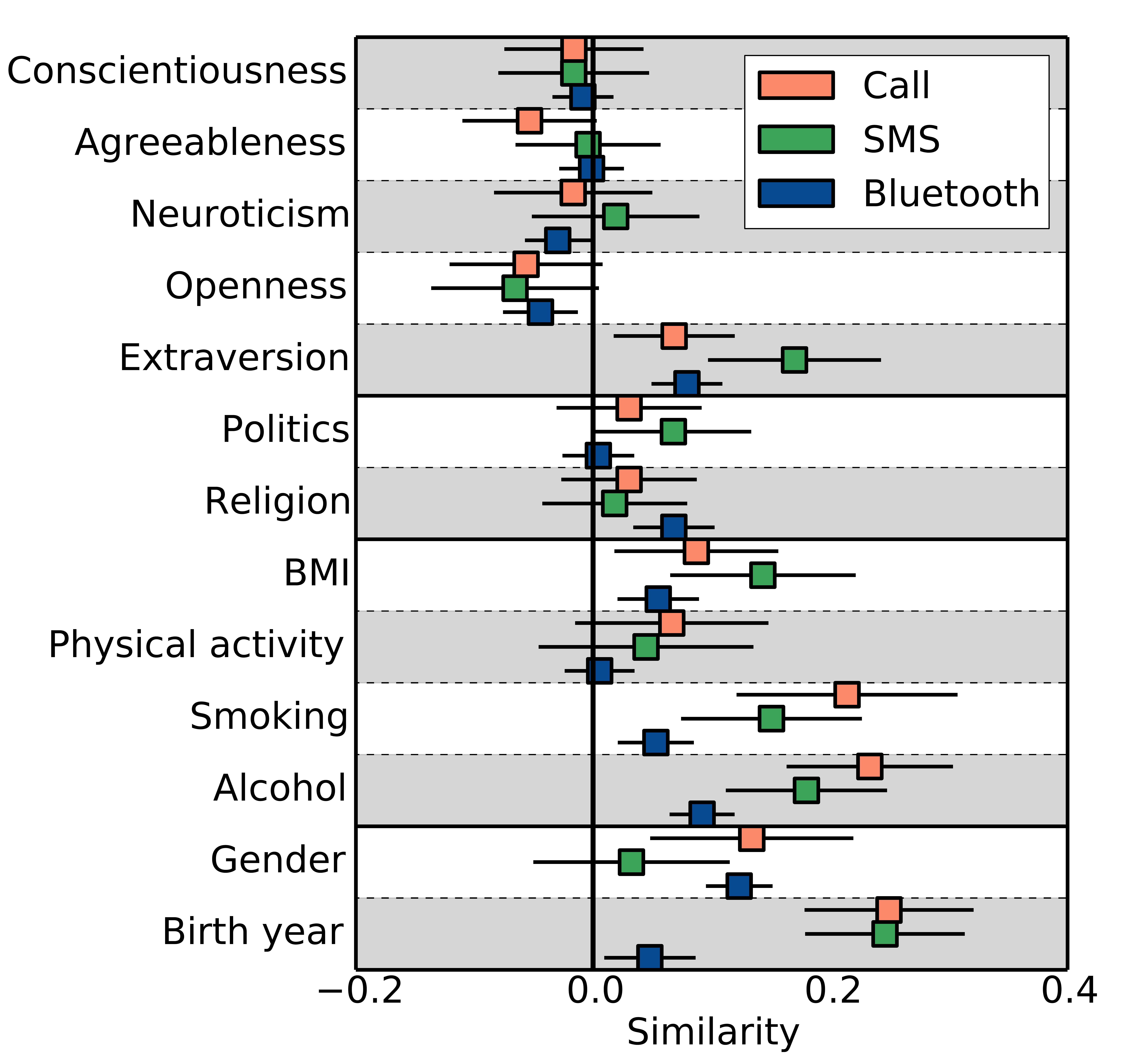}
	\caption{ {\bf Similarity of connected individuals in the network.} The bars show the intraclass correlation coefficient for the different variables and for the networks formed from call activity, SMS activity and proximity data. The lines have a range of one standard deviation. We find no similarity with regard to the personality traits conscientiousness, agreeableness and neuroticism but we find a weak similarity with regard to extraversion and dissimilarity with respect to openness. In general a stronger similarity is found for socio-demographic, behavior- and attitudes-related variables.\label{fig:Similarity}}
\end{figure}

Here, we consider three different types of social interaction networks based on calls, text messages, and physical proximity, respectively. We introduce a tunable link weight based on the strength of the interactions. To explain our definition of a link weight, let us start by considering the call network. The weight of a directed link from person $i$ to person $j$ is given by $$w_{ij}(\alpha)=n_{ij}^\alpha/\sum_k n_{ik}^\alpha,$$  where $n_{ij}$ represents the total number of accepted calls from person $i$ to person $j$. Links therefore take a value in the interval, $w_{ij}\in\left[0,1\right]$, and the sum of weights of outgoing links from any person equals unity. The power $\alpha$ is used to test if homophily is more pronounced between individuals who interact more frequently than for individuals who do not interact that often. The case $\alpha=0$ corresponds to a network where all links have equal weight. For intermediate $\alpha$ values, we predominantly test for similarity on the strongest links and, ultimately, for large values, e.g., for $\alpha\approx 2$, we only consider the strongest out-going link for each individual. The network of text-messages (SMS network) is constructed in the similar fashion, but with $n_{ij}$, representing the number of text messages sent from person $i$ to person $j$. From the data on physical proximity, we can determine the time a pair of individuals has spent together. We say that a person $i$ has spent an amount of time $\Delta t$ together with person $j$ if two consecutive Bluetooth scans are separated by a time $\Delta t$ and, in addition, both scans estimate person $j$ to be within approximately three meters distance. The link weight between $i$ and $j$ is $$w_{ij}(\alpha) = T_{ij}^\alpha / \sum_k T_{ik}^\alpha,$$ where $T_{ij}$ is the total time that $j$ has been within the three meter limit of $i$. In general, the proximity data contains information about a large number of more or less random encounters during lectures and classes. In order to prevent that these encounters dominate our data, we make use of proximity data sampled only in the weekends or from 6pm to 12am during the weekdays. We place no such restrictions on the call and SMS data. Finally, we construct a symmetric weight from the two directed weights by taking the average weight of the two directed links. From these three types of interaction, we construct the corresponding networks, see Fig.~\ref{fig:networks}. Here the size and color of the nodes are determined by the sum of link weights connecting to the node, while the width of a link is given by the square root of the link weight. The visual representation reveals that the networks tend to be dominated by a relatively small set of links with strong weights.

In order to analyze homophily, we construct vectors $(x_i,x_j,w_{ij})$ for each link in the network where $x_i$ represents a variable (e.g., of a personality trait) associated with person $i$. The degree of homophily is estimated by a generalization of the intraclass correlation coefficient (ICC). The ICC quantifies the similarity of the variables $x_i$ and $x_j$ for the connected persons $i$ and $j$ in the network. Similarly to the Pearson correlation coefficient, the ICC is a measure of the tendency for $x_i$ and $x_j$ to assume similar values relative to their average value. Normally, the ICC is computed under the assumption that persons are either connected or not. Here we modify the ICC by including the weight of interactions $w_{ij}$ between persons. The weighted ICC, here denoted by  $r$, is then computed for a network, $(x_i,x_j,w_{ij})$, from the expressions,

\begin{align*}
r &= t^2/s^2\\
\bar{x} &= \frac{\sum_{i>j} w_{ij} \left( x_i + x_j \right)}{\sum_{i>j} 2w_{ij}}, \\
s^2 &= \frac{\sum_{i>j} w_{ij} \left( \left( x_i - \bar{x} \right)^2 + \left( x_j - \bar{x} \right)^2	 \right)}{\sum_{i>j} 2w_{ij}}, \\
t^2 & = \frac{\sum_{i>j} w_{ij} \left( x_i - \bar{x} \right)  \left( x_j - \bar{x} \right)	}{\sum_{i>j} 2w_{ij}}, \\
\end{align*}
The auxiliary variable $s$ measures the variance within the sample, including both variables $x_i$ and $x_j$, and the variable $t$ is a measure of the co-variance of $x_i$ and $x_j$. Please note how the contribution to the variance for each link is weighted by $w_{ij}$. In general, the weighted correlation coefficient provides a basic measure of the importance of homophily in social interactions. In Fig.~\ref{fig:Similarity}, we show the ICC where all weights are proportional to the activity on the link, i.e., $\alpha=1$. The error bars are estimated using bootstrapping, where we for each value of $\alpha$ and for each network layer (Call, SMS, and BlueTooth), generate 10,000 reference networks by randomly reshuffling the links. We then measure the correlation coefficient in these reference network. The fraction of networks with an ICC larger than that of the true network provides us with a measure of the p-value. 

\begin{figure*}
	\centering
	\includegraphics[width=\textwidth]{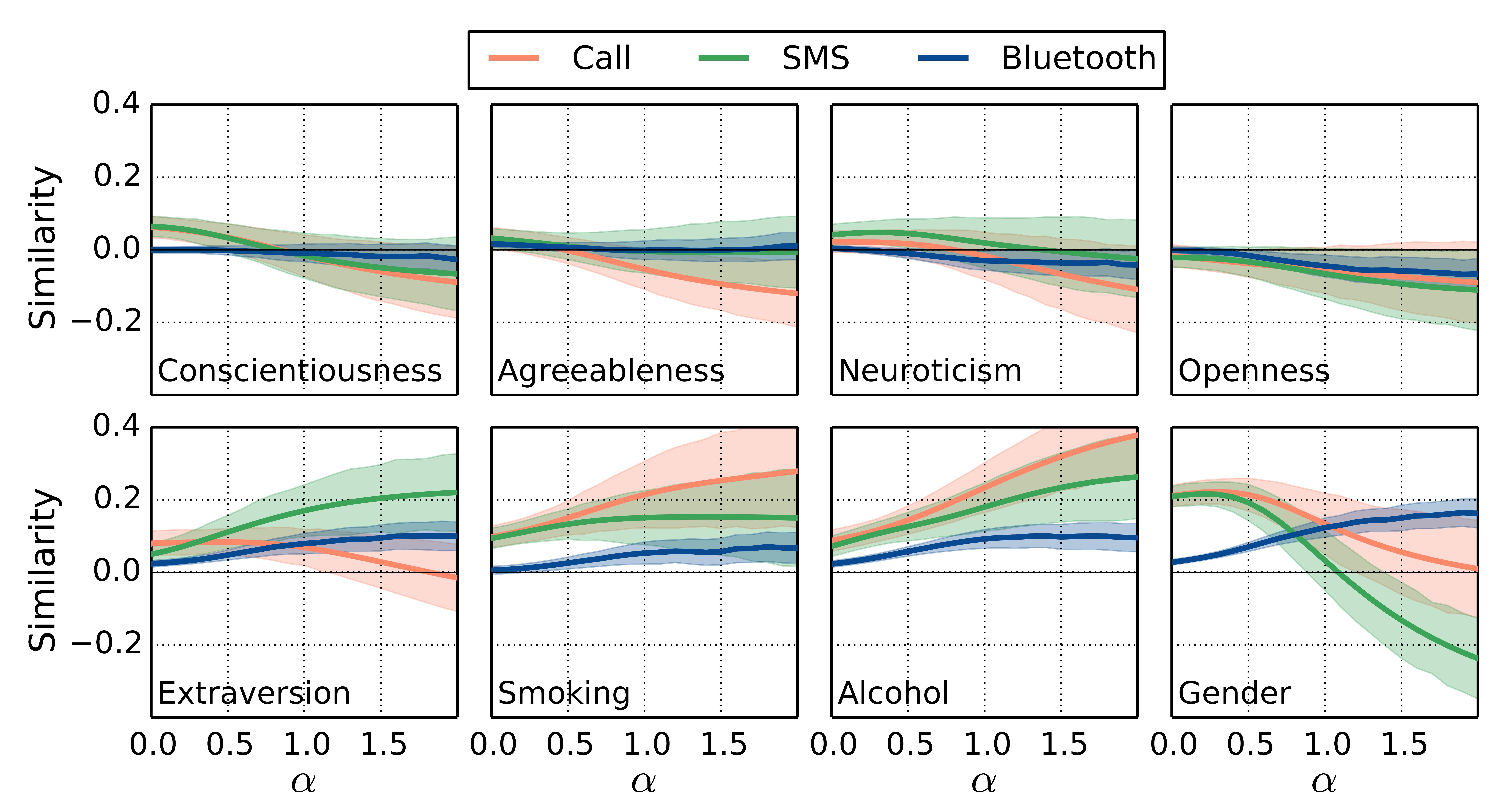}
	\caption{ {\bf Computed similarity of individuals with different levels of social interaction.} The computed similarity is shown as function of the parameter $\alpha$, which tunes the weight of the individual links in the networks. The case $\alpha=0$ corresponds to the case of links having equal weight, whereas increasing values of $\alpha$ enhances the contribution from the stronger links in the calculation, e.g., for $\alpha=2$ the strongest links of each individual dominates. The envelope is the estimate of the standard deviation (see text). In the Supporting Information S3, a table is included of p-values for the individual traits and for different $\alpha$-values. In general, the p-values are large ($>0.05$) for the traits Conscientiousness, Agreeableness, Neuroticism and Openness. The estimates of Extraversion attains lower p-values, e.g. for the BlueTooth, $p\approx 0.01$, and for SMS the p-value varies in the interval $0.02-0.08$ for different $\alpha$-values.   \label{fig:simalpha}}
	\end{figure*}

 We observe in Fig.~\ref{fig:Similarity} that there is no pronounced homophily for the personality traits conscientiousness, agreeableness and neuroticism, even when we consider only the strongest links.  In Fig.~\ref{fig:simalpha}, we test the importance of link strength by varying the parameter $\alpha$, i.e., we test for homophily by considering all social interactions equally important ($\alpha=0$) or by weighting frequent interactions higher ($\alpha>0$). We see that for the Big Five personality traits, only extraversion have ICCs which are significantly different ($p<0.05$) from zero in all layers. List of p-values for the computed ICCs are listed in the Supporting Information S3 and a description of how the p-values are computed can be found in Materials and Methods. In the sms layer, the ICC for extraversion ranges from values around zero when all links have equal weights to values around 0.2 for $\alpha=2$.  
For both the extraversion and openness traits, the proximity and call layers result in ICCs that are lower than the ICC of the text message layer. The ICCs for agreeableness, conscientiousness and neuroticism are for almost all values of $\alpha$ not significantly different from zero and are bounded above by approximately 0.12.

Homophily is pronounced in the phone call network for the variables capturing smoking and drinking behaviors. Here the ICCs are significantly different ($p<0.05$) from zero and achieve values larger than $0.3$ in the call layer and values up to 0.2 in the sms layer. This is in contrast to the other variables in our study, where homophily is most pronounced in the sms layer. The variables representing attitudes concerning politics and religion show a weak or no correlation. Less surprisingly, we see an over-representation of social interaction between individuals of the same sex for calls and text messages when $\alpha=0$; the ICCs attain values around $0.2$. Moreover, we observe that for increasing values of $\alpha$, the stronger links in the text message network more frequently connect individuals of different sex, i.e.\ we see a transition from a positive ICC to a negative ICC as alpha is increased. Interestingly, albeit the correlation is slightly smaller, the proximity data at the same time shows that individuals with frequent face-to-face encounters tend to be of same sex.

\section*{Discussion}

\begin{figure*}
	\centering
	\includegraphics[width=\textwidth]{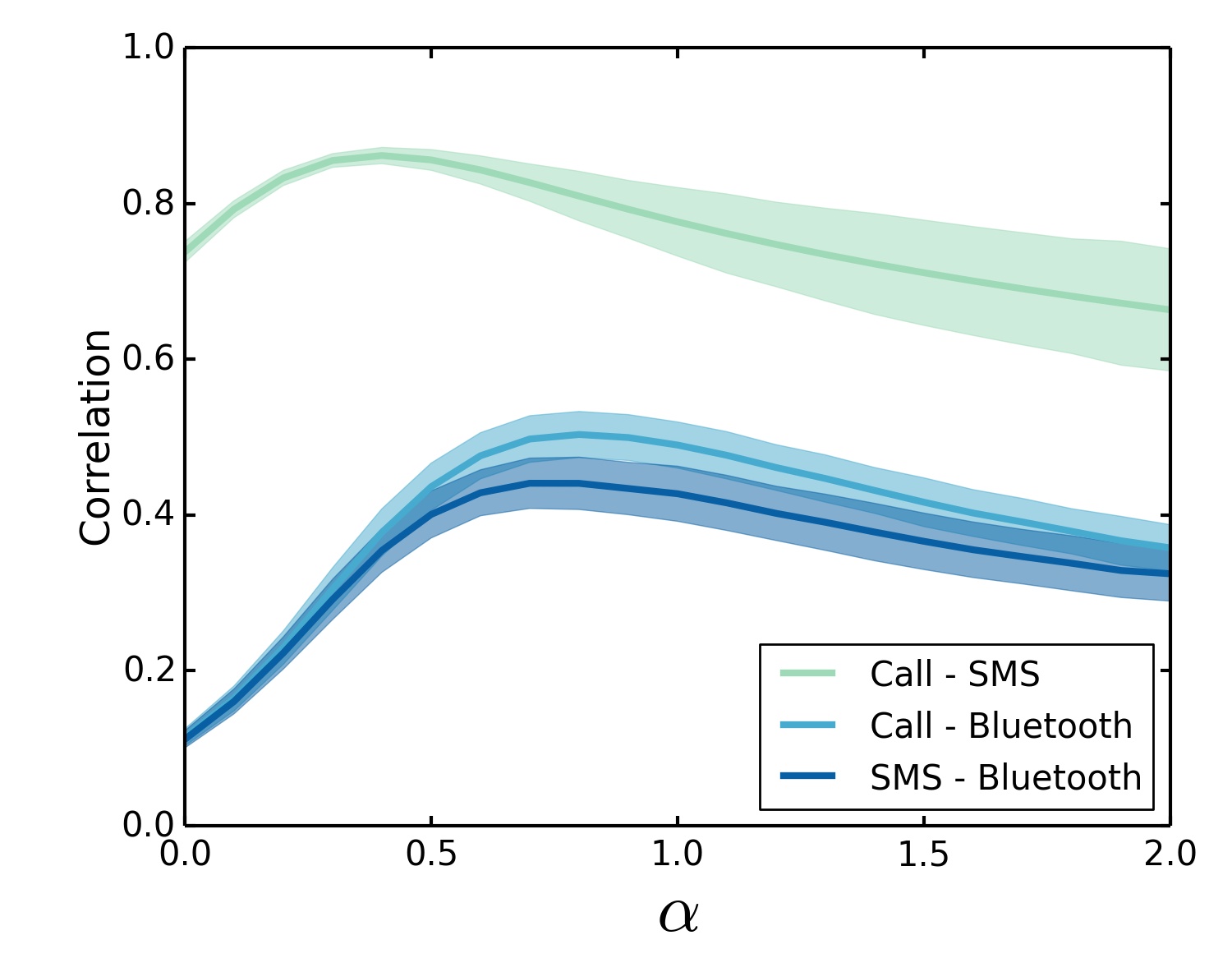}
	\caption{ {\bf Overlap between layers in the multilayer social network.} Based on the correlation between the link-weights, we can check the overlap between the individual layers of call, SMS, and proximity for varying values of $\alpha$. We see that the multilayer features of the network changes with the link-weight exponent $\alpha$ and the overlap between the layers is maximal for intermediate values of $\alpha$. \label{multilayer}}
\end{figure*}

{\bf Multilayer networks --} 
The overlap between the three layers in the multilayer network can be estimated from the pairwise Pearson correlation coefficients $r_{p,k\ell}$ of the link weights $w_{ij}^L$ in two layers $L_k$ and $L_\ell$.
\begin{equation}
r_{p,k\ell}=\frac{\sum_{i>j}(w^{L_k}_{ij}-\bar w^{L_k})(w^{L_\ell}_{ij}-\bar w^{L_\ell})}{\left[\sum_{i>j}(w^{L_k}_{ij}-\bar w^{L_k})^2\right]^{1/2}\left[\sum_{i>j}(w^{L_\ell}_{ij}-\bar w^{L_\ell})^2\right]^{1/2}}
\end{equation}
We find that for $\alpha=1$ the correlation coefficient is 0.75 between the call and SMS layers, 0.53 between the call and proximity layers, and 0.47 between the SMS and proximity layers. A similar approach has previously been suggested in Ref.~\cite{nicosia2015measuring} where, instead of the link weights, the degree of the nodes in the individual layers was considered. Using the link weights, we can now by tuning the parameter $\alpha$ test the overlap between the layeres for different levels of acquaintanceships. In Fig. \ref{multilayer}, we show the pairwise correlation between the three layers for different values of $\alpha$. As expected there is a significant overlap between the layers, but they certainly also differ enough to be treated as more than a fluctuation of a single network. Interestingly, the overlap changes with the factor $\alpha$, which opens a fundamental question in the analysis of multiplex networks. Which weights would be the right to use? The unweighted case $\alpha=0$ certainly leads to a correlation different than those of larger $\alpha$ values. In fact, strong links might not necessarily be present or strong in all layers, e.g. two persons that frequently communicate might prefer phone calls rather than text messages. At the intermediate range, interaction could be more equally distributed across the channels or layers. In other words, the degree of multiplexity in our network is tunable and depends on the perspective, whether strong or weak links should be favored. This observed sensitivity in  overlap, could have implications for community detection algorithms on multiplex networks \cite{Mucha876, PhysRevX.5.011027} or for the structural reducibility of overlapping layers\cite{de2015structural}.

We further note that the proximity (Bluetooth) layer is more densely connected than the other layers, in particular because the participants in the study meet at more informal gatherings at the university campus or have encounters which could either be spontaneous or of less personal character such as study groups. This could be one reason for the weaker similarity seen for most of the variables in the proximity data in Figs.~\ref{fig:Similarity} and ~\ref{fig:simalpha}.

Here, we have performed an extensive mapping of similarity in a large social network based on detailed records of social interactions over a time span of nearly two years. From the frequency of interactions, all links in the network are assigned a weight, which we have been able to tune in order to look for homophily across varying levels of acquaintanceships. We show that tuning the weights can reveal new features of the node similarity. For the variables describing alcohol use, cigarette use, and extraversion, we see that individuals are more similar when they interact strongly.  In contrast if the weights are disregarded, we see little or no similarity. Interestingly, the similarity of individuals is not monotonically increasing with the frequency of interaction for all variables, e.g., the intraclass correlation coefficient with regard to gender transitions from postive to negative values. The analysis of our data does not provide any evidence that the basic personality traits agreableness, conscientiousness, neuroticism and to some degree openness are an important factor in the formation of social networks. In fact, we find a small or non-existing correlation between these personality traits and social interaction, even when we only consider individuals that interact very frequently. Finally, the measure, we have introduced, shows that the degree of muliplexity in our network is tunable as we vary the balance between weak and strong links.

\section*{Materials and Methods}
In constructing the multilayer network, we include links from participants that meet minimum requirements with respect to the total time window in which they are active and their level of activity. In particular, we require that the data recording period is longer than 3 months and associated with at least 170 calls, 950 text messages and 200 hours of Bluetooth interaction. These numbers correspond to the typical social activity of a person during a 3 months period, which, we believe, is a reasonable time scale for the resolution of social behavior. These requirements reduce the dataset to 659 participants and is introduced to avoid the addition of noisy links in the network. The average user in the study has been active for 530 days, has been part of 952 phone calls, and has exchanged 5313 text messages. The average number of hours that a user has been in the proximity of others is 1073.  The proximity network is based on asynchronous Bluetooth scans by each smartphone every 5 minutes, which are collected into 5 minute time-bins and symmetrized.  Many of the recorded interactions are with people outside the study and can therefore not be included in the analysis of homophily. In the call and SMS data, the total weight of a single individual therefore depends on the fraction of calls or text messages that are directed to other participants in the study. 

The significance of our estimated ICCs have been computed in the following way. For each value of alpha and each layer in the network, we generate 10,000 reference layers (i.e.\ networks) by shuffling the links within a layer. We then measure the ICC in these reference layers. The fraction of network layers with an intraclass correlation coefficient larger than that of the original network layer provides us with a estimated of the p-value. A table of all computed p-values have been included in the Supporting Information S3.

This study was reviewed and approved by the appropriate Danish authority, the Danish Data Protection Agency (Reference number: 2012-41-0664).
The Data Protection Agency guarantees that the project abides by Danish law and also considers potential ethical implications. All subjects in the study provided written informed consent.

\bibliography{reference_plos}

\begin{thebibliography}{10}

\bibitem{borgatti2009network}
Borgatti SP, Mehra A, Brass DJ, Labianca G.
\newblock Network analysis in the social sciences.
\newblock science. 2009;323(5916):892--895.

\bibitem{eagle2006reality}
Eagle N, Pentland A.
\newblock Reality mining: sensing complex social systems.
\newblock Personal and ubiquitous computing. 2006;10(4):255--268.

\bibitem{10.1371/journal.pone.0011596}
Cattuto C, {Van den Broeck} W, Barrat A, Colizza V, Pinton J, Vespignani A.
\newblock Dynamics of Person-to-Person Interactions from Distributed RFID
  Sensor Networks.
\newblock PLOS ONE. 2010 07;5(7):e11596.
\newblock Available from:
  \url{http://dx.doi.org/10.1371%2Fjournal.pone.0011596}.

\bibitem{aharony2011social}
Aharony N, Pan W, Ip C, Khayal I, Pentland A.
\newblock Social fMRI: Investigating and shaping social mechanisms in the real
  world.
\newblock Pervasive and Mobile Computing. 2011;7(6):643--659.

\bibitem{10.1371/journal.pone.0095978}
Stopczynski A, Sekara V, Sapiezynski P, Cuttone A, Madsen MM, Larsen JE, et~al.
\newblock Measuring Large-Scale Social Networks with High Resolution.
\newblock PLoS ONE. 2014 04;9(4):e95978.
\newblock Available from: \url{http://dx.doi.org/10.1371/journal.pone.0095978}.

\bibitem{kivela2014multilayer}
Kivel{\"a} M, Arenas A, Barthelemy M, Gleeson JP, Moreno Y, Porter MA.
\newblock Multilayer networks.
\newblock Journal of Complex Networks. 2014;2(3):203--271.

\bibitem{Crandall1997133}
Crandall CS, Schiffhauer KL, Harvey R.
\newblock Friendship pair similarity as a measure of group value.
\newblock Group Dynamics. 1997;1(2):133--143.
\newblock Available from:
  \url{http://www.scopus.com/inward/record.url?eid=2-s2.0-0141662438&partnerID=40&md5=cfe73a3d2adb3e9e5b1eaed153451a2b}.

\bibitem{currarini2010identifying}
Currarini S, Jackson MO, Pin P.
\newblock Identifying the roles of race-based choice and chance in high school
  friendship network formation.
\newblock Proceedings of the National Academy of Sciences.
  2010;107(11):4857--4861.

\bibitem{Kandel1978306}
Kandel DB.
\newblock Similarity in real-life adolescent friendship pairs.
\newblock Journal of Personality and Social Psychology. 1978;36(3):306--312.
\newblock Available from:
  \url{http://www.scopus.com/inward/record.url?eid=2-s2.0-25444466854&partnerID=40&md5=00cc0153e389eaf6d02ccd89624333d1}.

\bibitem{Rushton2005555}
Rushton JP, Bons TA.
\newblock Mate Choice and Friendship in Twins: Evidence for Genetic Similarity.
\newblock Psychological Science. 2005;16(7):555--559.
\newblock Available from:
  \url{http://www.scopus.com/inward/record.url?eid=2-s2.0-23944492077&partnerID=40&md5=56a6b7b64363aae63c9f77e1478894cc}.

\bibitem{kurtz2003relationship}
Kurtz JE, Sherker JL.
\newblock Relationship quality, trait similarity, and self-other agreement on
  personality ratings in college roommates.
\newblock Journal of personality. 2003;71(1):21--48.

\bibitem{stehle2013gender}
Stehl{\'e} J, Charbonnier F, Picard T, Cattuto C, Barrat A.
\newblock Gender homophily from spatial behavior in a primary school: a
  sociometric study.
\newblock Social Networks. 2013;35(4):604--613.

\bibitem{Cohen2013816}
Cohen TR, Panter AT, Turan N, Morse L, Kim Y.
\newblock Agreement and similarity in self-other perceptions of moral
  character.
\newblock Journal of Research in Personality. 2013;47(6):816--830.
\newblock Available from:
  \url{http://www.scopus.com/inward/record.url?eid=2-s2.0-84884293708&partnerID=40&md5=c463cc29df8e4a105ca67e0dc8ae9af9}.

\bibitem{Lee2009460}
Lee K, Ashton MC, Pozzebon JA, Visser BA, Bourdage JS, Ogunfowora B.
\newblock Similarity and Assumed Similarity in Personality Reports of
  Well-Acquainted Persons.
\newblock Journal of Personality and Social Psychology. 2009;96(2):460--472.
\newblock Available from:
  \url{http://www.scopus.com/inward/record.url?eid=2-s2.0-60749103381&partnerID=40&md5=8ba442558d8e4f4298385a93f3ccb448}.

\bibitem{Paunonen2013800}
Paunonen SV, Hong RY.
\newblock The many faces of assumed similarity in perceptions of personality.
\newblock Journal of Research in Personality. 2013;47(6):800--815.
\newblock Available from:
  \url{http://www.scopus.com/inward/record.url?eid=2-s2.0-84884223362&partnerID=40&md5=f33fe27c11db6a1b460225611fd29858}.

\bibitem{Delay2013464}
Delay D, Laursen B, Kiuru N, Salmela-Aro K, Nurmi JE.
\newblock Selecting and retaining friends on the basis of cigarette smoking
  similarity.
\newblock Journal of Research on Adolescence. 2013;23(3):464--473.
\newblock Available from:
  \url{http://www.scopus.com/inward/record.url?eid=2-s2.0-84882664852&partnerID=40&md5=37a95e3589dc5e9d318e06c0249d6737}.

\bibitem{john2008paradigm}
John OP, Naumann LP, Soto CJ.
\newblock Paradigm shift to the integrative big five trait taxonomy.
\newblock Handbook of personality: Theory and research. 2008;3:114--158.

\bibitem{mollgaard2016dynamics}
Mollgaard A, Mathiesen J.
\newblock The Dynamics of Initiative in Communication Networks.
\newblock PloS one. 2016;11(4):e0154442.

\bibitem{sekara2014strength}
Sekara V, Lehmann S.
\newblock The strength of friendship ties in proximity sensor data.
\newblock PLOS One. 2014;9(7).

\bibitem{rosenberg1989society}
Rosenberg M.
\newblock Society and the adolescent self-image (rev).
\newblock Wesleyan University Press; 1989.

\bibitem{back2013narcissistic}
Back MD, K{\"u}fner AC, Dufner M, Gerlach TM, Rauthmann JF, Denissen JJ.
\newblock Narcissistic admiration and rivalry: Disentangling the bright and
  dark sides of narcissism.
\newblock Journal of Personality and Social Psychology. 2013;105(6):1013.

\bibitem{diener1985satisfaction}
Diener E, Emmons RA, Larsen RJ, Griffin S.
\newblock The satisfaction with life scale.
\newblock Journal of personality assessment. 1985;49(1):71--75.

\bibitem{rotter1966generalized}
Rotter JB.
\newblock Generalized expectancies for internal versus external control of
  reinforcement.
\newblock Psychological monographs: General and applied. 1966;80(1):1.

\bibitem{russell1996ucla}
Russell DW.
\newblock {UCLA} Loneliness Scale (Version 3): Reliability, validity, and
  factor structure.
\newblock Journal of personality assessment. 1996;66(1):20--40.

\bibitem{sherer1982self}
Sherer M, Maddux JE, Mercandante B, Prentice-Dunn S, Jacobs B, Rogers RW.
\newblock The self-efficacy scale: Construction and validation.
\newblock Psychological reports. 1982;51(2):663--671.

\bibitem{cohen1983global}
Cohen S, Kamarck T, Mermelstein R.
\newblock A global measure of perceived stress.
\newblock Journal of health and social behavior. 1983;p. 385--396.

\bibitem{bech2001sensitivity}
Bech P, Rasmussen NA, Olsen LR, Noerholm V, Abildgaard W.
\newblock The sensitivity and specificity of the Major Depression Inventory,
  using the Present State Examination as the index of diagnostic validity.
\newblock Journal of affective disorders. 2001;66(2):159--164.

\bibitem{lund2014content}
Lund R, Nielsen LS, Henriksen PW, Schmidt L, Avlund K, Christensen U.
\newblock Content Validity and Reliability of the {Copenhagen} {Social}
  {Relations} {Questionnaire}.
\newblock Journal of aging and health. 2014;26(1):128--150.

\bibitem{watson1988development}
Watson D, Clark LA, Tellegen A.
\newblock Development and validation of brief measures of positive and negative
  affect: the {PANAS} scales.
\newblock Journal of personality and social psychology. 1988;54(6):1063.

\bibitem{nicosia2015measuring}
Nicosia V, Latora V.
\newblock Measuring and modeling correlations in multiplex networks.
\newblock Physical Review E. 2015;92(3):032805.

\bibitem{Mucha876}
Mucha PJ, Richardson T, Macon K, Porter MA, Onnela JP.
\newblock Community Structure in Time-Dependent, Multiscale, and Multiplex
  Networks.
\newblock Science. 2010;328(5980):876--878.
\newblock Available from:
  \url{http://science.sciencemag.org/content/328/5980/876}.

\bibitem{PhysRevX.5.011027}
De~Domenico M, Lancichinetti A, Arenas A, Rosvall M.
\newblock Identifying Modular Flows on Multilayer Networks Reveals Highly
  Overlapping Organization in Interconnected Systems.
\newblock Phys Rev X. 2015 Mar;5:011027.
\newblock Available from:
  \url{http://link.aps.org/doi/10.1103/PhysRevX.5.011027}.

\bibitem{de2015structural}
De~Domenico M, Nicosia V, Arenas A, Latora V.
\newblock Structural reducibility of multilayer networks.
\newblock Nature communications. 2015;6.

\end{thebibliography}

\section*{Acknowledgements }
The study received funding through the UCPH 2016 Excellence Programme for Interdisciplinary Research. 
\section*{Supporting Information} 
\subsection*{S1 Text} Description of data. Short description of the format of the data.
\subsection*{S2 Text} Data file. Data used in the analysis in the manuscript.
\subsection*{S3 Text} List of p-values for different traits and values of $\alpha$.

\end{document}